\documentclass[prl,twocolumn,showpacs,preprintnumbers,amsmath,amssymb]{revtex4}

\usepackage{graphicx}
\usepackage{dcolumn}
\usepackage{bm}

\begin{document}

\preprint{APS/123-QED}

\title{Polarization effects in attosecond photoelectron spectroscopy}

\author{Jan Conrad Baggesen}
 \affiliation{Lundbeck Foundation Theoretical Center for Quantum System Research \\
Department of Physics and Astronomy, Aarhus University, 8000 Aarhus C, Denmark.}
 
\author{Lars Bojer Madsen}%
 \affiliation{Lundbeck Foundation Theoretical Center for Quantum System Research \\
Department of Physics and Astronomy, Aarhus University, 8000 Aarhus C, Denmark.}

\date{\today}

\begin{abstract} 
We study the influence of polarization effects in streaking by combined atto- and femtosecond pulses. The polarization-induced terms alter the streaking spectrum. The normal streaking spectrum, which maps to the vector potential of the femtosecond pulse, is modified by a contribution following the field instead. 
We show that polarization effects may lead to an apparent temporal shift, that needs to be properly accounted for in the analysis. The effect may be isolated and studied by angle-resolved photoelectron spectroscopy from oriented polar molecules. We also show that polarization effects will lead to an apparent temporal shift of 50 as between photoelectrons from a 2p and 1s state in atomic hydrogen. 
\end{abstract}

\maketitle

\begin{figure}[h!]
	\includegraphics[width = 0.40\textwidth]{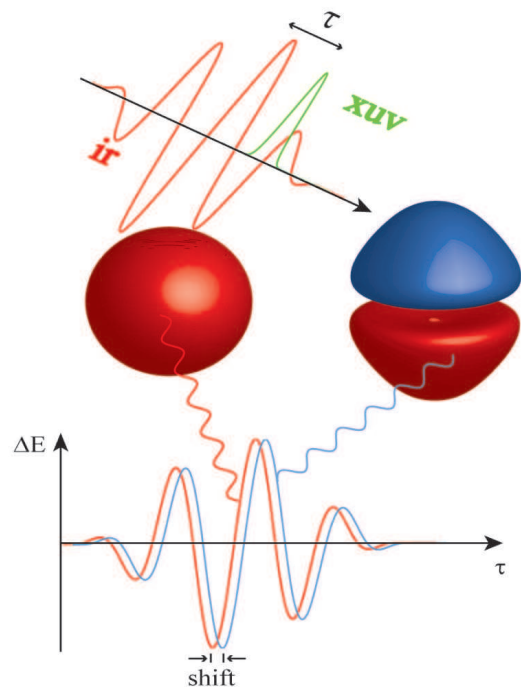}
	\caption{(Color online) Schematic presentation of the attosecond streaking experiment. A fs ir pulse and an attosecond xuv pulse with a variable delay, $\tau$, are focused onto a target. The kinetic energy of the photoelectrons depends on the delay between the two pulses, which gives a characteristic signal. If electrons are released from two different channels (sketched as the s-like and the p-like orbitals) and if release from one of the channels  is delayed compared to the other, the signal is shifted as shown in the bottom of the figure. See~\cite{cavalieri2007} for an experimental realization.}
	\label{fig:streaking}
\end{figure}

The attosecond (1 as = 10$^{-18}$ s) defines the natural time scale for electronic motion within atoms and molecules, just as the femtosecond (fs) is the natural time scale for nuclear motion in molecules. This is one reason for the large current interest in attosecond science~\cite{krausz2009}. Photoelectron spectroscopy with combined extreme ultraviolet attosecond (xuv) and few-cycle near-infrared (ir) pulses is used both to characterize attosecond pulses~\cite{itatani2002, kienberger2004}, few-cycle laser pulses~\cite{goulielmakis2004} and to measure ultrafast electron dynamics~\cite{uiberacker2007,cavalieri2007,miaja-avila2008,mauritsson2008}.
Attosecond streaking is a very promising tool for time-resolved measurements with sub-fs resolution. In attosecond streaking, one exploits that the attosecond xuv pulse is very short compared to the optical period of the assisting ir field. Then, the electrons released by the xuv pulse are all released at a definite phase of the ir pulse and obtain a momentum change due to the propagation in the ir field given classically by $\Delta \vec k(\tau) = -\int_\tau^\infty \vec F(t)dt = -\vec A(\tau)$, $\tau$ is the time of ionization by the attosecond pulse, $\vec{F}(t)$ is the electric field and $\vec A(t)$ the vector potential [atomic units (a.u.) with  $\hbar = e = a_0 = m_e = 1$ are used throughout unless indicated otherwise]. To obtain a clear streaking spectrum, the ir pulse should be sufficiently intense that the streaking momentum change is clearly seen, but still sufficiently weak not to excite or ionize the target. A schematic presentation of a streaking experiment is shown in Fig.~\ref{fig:streaking}.

In this work, we show that if the target is polarized by the ir pulse, the conventional streaking spectrum, mapping to the vector potential of the ir field, is modified by a dipole term proportional to the field and a polarizability term quadratic in the field. 
Two different curves are plotted in the bottom of Fig.~\ref{fig:streaking}, corresponding to two different signals, in this case electrons released from different orbitals. The shift between the two curves may be interpreted as a temporal delay due to difference in emission time between the two signals.
The linear shift following $F$ is out of phase with the vector potential and when  the phase of the streaking signal is used as a measure of emission time, such a shift will manifest itself as an apparent time-delay, that needs to be properly accounted for in the analysis. 
We show that apparent time-delays may be suppressed by using attosecond pulses with shorter wavelengths.

We model the electron emission from an $N$-electron 
atom or molecule in a $T$-matrix formalism~\cite{bojer2005}, with the differential probability for emission with momentum 
$\vec k_f$ given by $dP/d\vec k_f = \left| T_{fi} \right|^2$ and
$T_{fi} = -i\int^\infty_{-\infty} \langle \Psi_f(t) | V_x(t) | \Psi_i(t)\rangle dt$, where  $V_x(t) = \sum_{j=1}^{N}\vec r_j \cdot \vec F_x(t)$ describes the interaction of the electrons with the xuv light. To model the polarization effect, we assume for simplicity that the target system has a permanent dipole moment, $\vec\mu$. Hence, we include the Stark shift due to the relatively weak ir pulse through first-order perturbation theory and the initial state is given by $|\Psi_i(t)\rangle = \exp\left[-i \int^t(E_i - \vec \mu\cdot \vec F(t'))dt'\right]|\Psi_i\rangle$, with $|\Psi_i\rangle = \frac{1}{\sqrt{N!}}\text{det}\left|\psi_1(\vec r_1)\dots\psi_N(\vec r_N)\right|$ a Slater determinant of single-particle states. 

For the final state, we refer to the short duration of the attosecond pulse and assume an unrelaxed electronic state, with $N-1$ electrons in the same states as in $|\Psi_i\rangle$ and one single-particle state replaced by a Volkov wavefunction, describing a free particle in an electromagnetic field, 
$|\Psi_f(t)\rangle = \exp \left[-i \int^t(E_f - \vec \mu_{\text{ion}} \cdot \vec F(t'))dt' \right]  \vert \Psi_f \rangle$, 
with 
$ |\Psi_f\rangle = \frac{1}{\sqrt{N!}}\text{det}\left|\psi_1(\vec r_1)\dots\psi_{N-1}(\vec r_{N-1})\psi^V_{\vec k_f}(\vec r_N,t)\right|$, 
$\psi^V_{\vec k_f}(\vec r_N,t) = (2\pi)^{-3/2} e^{i(\vec k_f + \vec A(t))\cdot \vec r -\frac{i}{2}\int^t(\vec k_f + \vec A(t'))^2dt'}$, and $\vec A(t)$ the vector potential, $\vec F(t) = - \partial_t \vec A(t)$. 
The many-electron $T$-matrix element then reduces to a one-electron matrix element, 
$T_{fi} = -i\int_{-\infty}^\infty \langle\psi^V_{\vec k_f}(t) | {\vec r \cdot \vec F_x(t)} | \psi_i(t)\rangle e^{i\int^t( I_p + \Delta\vec\mu\cdot \vec F(t'))dt'} dt$, where $I_p = E_f - E_i$ is the ionization potential of the active orbital and $\Delta\vec\mu = \vec\mu - \vec\mu_{\text{ion}}$ is the difference in the dipole moments between the neutral and the unrelaxed ion.

Introducing the envelope function for the xuv pulse and maintaining only the absorption term, one may write $\vec F_x(t) = F_{x,0}\vec\epsilon_x  e^{-i\omega_x t}f_x(t-\tau)$, with $F_{x,0}$ the peak field strength, $\vec\epsilon_x$ the polarization vector, $\omega_x$ the angular frequency, $\tau$ the delay of the attosecond pulse to the ir pulse, and $f_x(t) = \exp(-t^2/T_x^2)$ a Gaussian envelope, with full-width-half-maximum, $T_x^{\text{FWHM}} = \sqrt{2\log(2)}T_x$. Collecting terms, we obtain
\begin{equation}
	T_{fi}(\tau) = -i\int_{-\infty}^{\infty}T^\text{1B}(\vec k + \vec A(t))e^{i\Phi(t)}f_x(t-\tau)dt,
	\label{eq:T_tau}
\end{equation}
where $T^\text{1B}(\vec k) = \langle \vec k|{F_0\vec\epsilon_x\cdot \vec r}|\psi_i\rangle$ is the first Born transition matrix element from the initial orbital to a free-electron final state with momentum $\vec k$ and 
\begin{equation}
	\Phi(t) = I_pt - \omega_x t + \frac{1}{2}\int^t\left[(\vec k + \vec A(t'))^2 + \Delta\vec\mu\cdot\vec F(t') \right]dt'
	\label{eq:Phi}
\end{equation} contains the phase variation of the integrand. Now, assuming the duration of the xuv pulse is very short compared to the time scale of the variation in the ir field, $\Phi(t)$ may be Taylor expanded to first order in $t-\tau$ as $\Phi(t) \sim \Phi(\tau) + (t-\tau)\frac{d\Phi}{dt}\big|_{t=\tau}$. With this approximation, and assuming $T^{1B}(\vec k + \vec A(t))$ is slowly varying with time, the time integration in the $T$-matrix element may be calculated analytically, giving
\begin{equation}
	\frac{dP}{d\vec k} = \left|T_{fi}^\text{1B}(\vec k + \vec A(\tau))\right|^2\pi T_x^2\exp\left(-\frac{T_x^2}{2}\frac{d\Phi}{dt}\Big|_{t=\tau}^2\right).
	\label{eq:dPdk}
\end{equation}
For each value of $\tau$, the electron energy distribution is centered around the electron energy 
\begin{equation}
	E(\tau) = \omega_x - I_p - \vec k\cdot \vec A(\tau) -  \Delta\vec\mu\cdot \vec F(\tau),
	\label{eq:Etau}
\end{equation}
which corresponds to $\frac{d\Phi}{dt}=0$. The $A(\tau)^2$-term of the ir pulse is  normally very small at the intensities used in streaking and has been neglected.  The streaking spectrum varies around the center energy as $- \vec k\cdot \vec A(\tau) -  \Delta\vec\mu\cdot \vec F(\tau)$, and is hence modified by the permanent dipole. If the intensity of the assisting ir field is increased the $ A(\tau)^2$-term and  the second-order Stark shift should be accounted for. In the quasi-static approximation, the center energy in the streaking spectrum is
\begin{equation}
\label{eq:E2}
E(\tau) =  \omega_x - I_p - \vec k\cdot \vec A(\tau) -  \frac{A(\tau)^2}{2} -\Delta\vec\mu\cdot \vec F(\tau) - \frac{1}{2} \vec F^\dagger ( \Delta {\bm \alpha}) \vec F, 
\end{equation}
where $\Delta {\bm \alpha}$ is the change in the polarizability between the neutral and the ion. 

Usually, it is assumed that the electron energy, plotted versus the delay between the two pulses, $\tau$, would follow the vector potential as $\vec k \cdot \vec A(\tau)$.
Polarization changes this result according to \eqref{eq:Etau}-\eqref{eq:E2}.
When using the $E(\tau)$ signal as the probe for time-resolved measurements, the $\vec F$-term gives rise to an apparent temporal shift in the streaking signal since it is out of phase with $\vec A$:
Neglecting the envelope function or looking at the peak of it, the two contributions from \eqref{eq:Etau} to the electron energy are $90^\circ$ out of phase and may be assumed to have the shape of $(\vec k \cdot \vec\epsilon_\text{ir}  F_0/\omega_\text{ir}) \sin(\omega_\text{ir}\tau)$ and $\Delta\mu\cdot \vec \epsilon_\text{ir}F_0\cos(\omega_\text{ir}\tau)$, respectively, where $A_0 = F_0/{\omega_\text{ir}}$ is the peak amplitude of the vector potential. Looking for the zero-crossings of this modulation with $\tau$, one finds that 
\begin{equation}
	\tan(\omega_\text{ir}\tau_d) = \omega_\text{ir} \frac{\Delta \vec{\mu}\cdot \vec\epsilon_\text{ir} }{\vec k \cdot \vec\epsilon_\text{ir}},	
	\label{eq:taud}
\end{equation}
where $\tau_d$ is the temporal change in the zero-crossings originating from including the Stark shift  in the calculation. This temporal shift depends on the angle of electron emission through the denominator. The second-order polarization effect \eqref{eq:E2} only affects the shape of the streaking spectrum, not the position of the peaks: An energy shift proportional to $F^2$ has a periodicity with twice the frequency of the vector potential. Adding such a contribution to the streaking signal will change the zero-crossings in both directions and will not change the positions of the peaks in the streaking spectrum.

\begin{figure}[]
	\includegraphics[width = 0.45\textwidth]{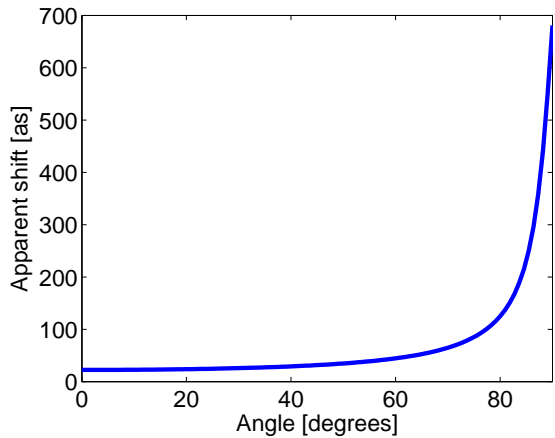}
	\caption{The apparent temporal shift versus the angle between $\vec k$ and $\vec \epsilon_\text{ir}$ for a polar molecule (see text). At $90^\circ$, the streaking spectrum follows the electric field instead of the vector potential (see \eqref{eq:Etau}).}
	\label{fig:angle}
	\end{figure}

The effect of the Stark shift on the streaking spectra is naturally explored in a polar molecule with a large permanent dipole moment. It is experimentally possible to align and orient polar molecules~\cite{holmegaard2009,de2009}, allowing a laboratory-fixed direction of the molecular dipole moment. Further, even in diatomic molecules the dipole moments may be rather large, making the apparent temporal shift more easily measureable.
As can be seen from \eqref{eq:taud}, the time-delay, $\tau_d$, of the photoelectrons from an oriented polar molecule should show a dependence on the angle between $\vec k$ and $\vec \epsilon_\text{ir}$. The apparent temporal shift of the photoelectrons against the detection angle is plotted in Fig.~\ref{fig:angle} in the case when the ir pulse is polarized parallel to the alignment of the molecule. We use the change in the dipole moment  of RbI, which is calculated to be $5.9\text{ D}$~\cite{gamess}. We use an electron energy of 85 eV and an ir wavelength of 800 nm. In the limit of $\vec\epsilon_\text{ir}\cdot\vec k \to 0$, corresponding to an observation orthogonal to the polarization axis, the apparent shift goes to $\frac{\pi}{2\omega_\text{ir}} = 680\text{ as}$, while the shift is only 22.3 as in the parallel geometry. This is because the streaking spectrum in the former case follows the electric field, $F$, instead of the vector potential, $A$, see \eqref{eq:Etau}. Since the strength of the electric field $F_\text{ir} = \omega_\text{ir}A_\text{ir}$ is much smaller than the vector potential, the amplitude of the oscillations is a factor of $\omega_\text{ir} \Delta \mu/k$ smaller in the perpendicular geometry.

In order to observe the shifts, we need to have an unshifted streaking spectrum for reference. We propose to mix-in ground-state He atoms, which are unpolarizable and have an ionization potential of 24.6 eV. The large ionization potential allows the separation of the photoelectrons emitted from He from those emitted from the highest occupied molecular orbital. 

Even though the temporal shift is most naturally studied in oriented molecules, for some molecules it might survive rotational averaging and show up from randomly oriented molecules. In a polar molecule, the intrinsic asymmetric electron distribution can cause the  ionization contribution from one orientation of the molecule to dominate the contributions from other orientations, leading to an apparent time-delay.

The polarization effects does not only occur in polar molecules. As another example, we calculate the streaking spectrum from a polarized $n=2$ state in atomic hydrogen. The Stark eigenstates~\cite{BransdenJoachain} of the $n=2$ states are given as $\psi_\pm = (\psi_{200} \pm \psi_{210})/\sqrt{2}$. These states have permanent dipole moments, $\vec\mu_\pm = \mp 3\hat z$. The H(1s) groundstate is spherical and non-degenerate, and may serve as a reference state, to show the shift. 

If H is initially in the experimentally accessible $\psi_{210}(t=-\infty) = (\psi_+ - \psi_-)/\sqrt{2}$ state, then the time-evolution of the two Stark eigenstates is simple, and at any later time, the system is in the state $\psi(t) = (\psi_+e^{i\Phi_+(t)} - \psi_-e^{i\Phi_-(t)})/\sqrt{2}$, where the subscripts refer to $\vec\mu_\pm = \mp 3\hat z$, see \eqref{eq:Phi}. The ionization probability in the positive $z$-direction is different for the two Stark eigenstates, and this results in a temporal shift.
\begin{figure}[]
	\includegraphics[width = 0.45\textwidth]{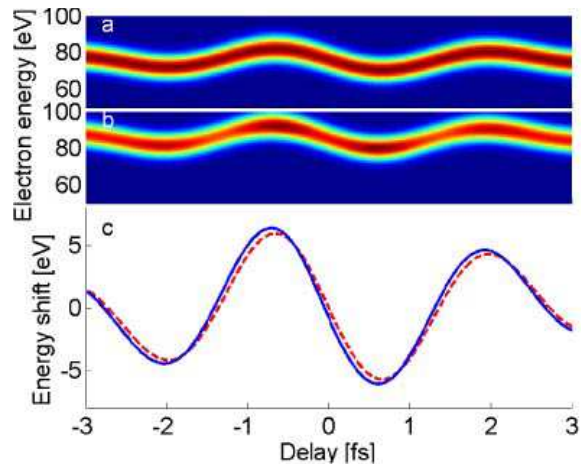}

	\caption{(Color online) (a) Streaking spectra from H(1s) and (b) from the $\psi(t)$-state in the positive $z$-direction (see text). The pulse parameters are presented in the text. (c) Center of energy analysis of the two spectra, showing that the signal from H(1s)  (full, blue) is shifted by 50 as compared to the signal from the $\psi(t)$-state (dashed, red).}
	\label{fig:210}	
\end{figure}
The effect shows up, since the electrons that are detected most often come from the $\psi_+$-state, as this state is more likely to emit electrons upwards, while the $\psi_-$-state is more likely to emit  downwards. The calculated streaking spectrum from $\psi(t)$ is shown in Fig.~\ref{fig:210}, and compared to the emitted spectrum from H(1s). The laser parameters used in our calculations are $T_x^\text{FWHM} = 290$ as, $\omega_x = 91$ eV and 800 nm ir pulse with $T_\text{ir} = 11$ fs and $I_\text{ir} = 10^{12}$ W/cm$^2$. Both the ir and the xuv pulse are polarized along the $z$-axis and the detector is placed in the direction $\hat z$ as well.

Figure \ref{fig:210} shows that the streaking signal from the excited state is "delayed" by 50 as compared to the groundstate. 
We conclude that the polarization induced apparent temporal shifts are important even when making measurements from un-polarized atoms, because of the selection in the ionization that is enforced through the angular resolved detection. Hence, this kind of effect may be important even in the ionization from spherical atoms, due to the fact that leaving the ion with one polarization may be more likely than leaving the ion with any other polarization.

In the bottom part of Fig.~\ref{fig:210}, a center of energy (coe) analysis of the streaking spectrum is performed. The center of energy is calculated as $\Delta E_\text{coe}(\tau) = \frac{\int EP(E,\tau)dE}{\int P(E,\tau)dE} - E_\text{coe}(\tau = -\infty)$, where $P(E,\tau)$ is the probability of emitting an electron with energy $E$, when the delay between the two pulses is $\tau$. 
The coe spectrum clearly displays the temporal shift.

In solids and particularly in metals, similar polarization effects may appear and give rise to modified streaking spectra \eqref{eq:Etau}-\eqref{eq:E2}. 
If the response of the electrons in the metal is sufficiently fast that there exist effects that follow the electric field on a few-fs time-scale, then any energy-shift proportional to the electric field would manifest itself as an apparent temporal shift, similar to the atomic and molecular cases discussed above. The complexity of the condensed phase many-electron system makes it difficult to make any accurate calculations, but similar effects might be at play and affect, e.g., the existing interpretations of the  measured time-delay  between photoelectrons from the 4f core and conduction band in tungsten~\cite{cavalieri2007} in terms of (i) difference in travel time~\cite{cavalieri2007,kazansky2009} or (ii) inter-layer interference~\cite{zhang2009}.  The theoretical interpretations of this experiment even hold quite different assumptions as to how the ir pulse penetrate the surface and show that the origin of the time-delay is not well understood~\cite{zhang2009,kazansky2009,lemell2009,baggesen2009}. 
Other surface experiments used the sidebands in the laser-assisted photoelectric effect~\cite{miaja-avila2008} to obtain temporal resolution on the scale of the duration of the ir pulse. This technique is sensitive only to the envelope and not the phase of the electric field, and polarization effects will accordingly affect the shape of the spectra but no apparent delays occur.
 
In conclusion, we have shown that polarization effects lead to modified streaking spectra \eqref{eq:Etau}-\eqref{eq:E2}, and that polarization may lead to apparent temporal shifts in attosecond streaking measurements, that needs to be properly accounted for. Attosecond streaking is a technique that shows very promising applications due to the sub-cycle resolution. As shown here the spectrum generally reflects a weighted sum of the vector potential and the electric field of the assisting fs ir pulse. The polarization terms may lead to  a shift in the streaking spectrum that does not necessarily reflect a delay in the photoelectron emission of one channel compared to another. This adds an extra complication to the interpretation of the experiments, as even a non-polar system as the hydrogen $\psi_{210}$-state may show this kind of apparent temporal shift. We have found an analytical expression~\eqref{eq:taud} for the apparent shift, showing that it can be avoided by going to higher electron energies, i.e., higher xuv photon energies. Further, we have proposed experiments to study the predicted effects in a controllable  way by angle-resolved photoelectron spectroscopy from oriented polar molecules.

This work was supported by the Danish Research Agency (Grant no. 2117-05-0081). We thank M. Abu-Samha for calculating the dipole moments of RbI and U. V. Poulsen for useful comments.


\end{document}